\documentclass[pra,longbibliography]{revtex4-1}
\usepackage{amsmath,amssymb}
\usepackage{epsfig}
\usepackage{dcolumn}
\usepackage{bbold}
\usepackage{natbib}
\usepackage{amsthm}
\usepackage{braket}
\usepackage{verbatim}
\usepackage{mdframed}
\usepackage[english]{babel}
\usepackage{bm}
\usepackage{graphicx}
\usepackage{epstopdf}
\usepackage{wrapfig}
\usepackage{array} 
\usepackage{listings}
\usepackage[para,online,flushleft]{threeparttablex}
\usepackage{booktabs,dcolumn}
\usepackage{color}
\newcolumntype{C}[1]{>{\centering\arraybackslash}p{#1}}
\usepackage{textpos}
\usepackage{booktabs}
\usepackage{makecell}
\usepackage{multirow,bigdelim}
\usepackage{float}
\usepackage{upgreek} 
\usepackage[utf8]{inputenc}
\usepackage{hyperref}
\hypersetup{breaklinks=true,colorlinks=true,linkcolor=blue,citecolor=blue,filecolor=magenta,urlcolor=blue}
\usepackage{xcolor}
\newcommand{\bs}[1]{\boldsymbol{{#1}}}

\begin{document}

\title{Efficient Solutions of Fermionic Systems using Artificial Neural Networks}

\author{Even M. Nordhagen}
\affiliation{Department of Physics and Njord Center, University of Oslo, N-0316 Oslo, Norway}

\author{Jane M. Kim}
\affiliation{Department of Physics and Astronomy and Facility for Rare Isotope Beams, Michigan State University, East Lansing, MI 48824, USA}

\author{Bryce Fore}
\affiliation{Physics Division, Argonne National Laboratory, Argonne, IL 60439, USA}

\author{Alessandro Lovato}
\affiliation{Physics Division, Argonne National Laboratory, Argonne, IL 60439, USA}

\author{Morten Hjorth-Jensen}
\affiliation{Department of Physics and Astronomy and Facility for Rare Isotope Beams, Michigan State University, East Lansing, MI 48824, USA}
\affiliation{Department of Physics and Center for Computing in Science Education, University of Oslo, N-0316 Oslo, Norway}

\begin{abstract}
{
We discuss differences and similarities between variational Monte Carlo approaches that use conventional and artificial neural network parameterizations of the ground-state wave function for  systems of fermions. We focus on a relatively shallow neural-network architectures, the so called restricted Boltzmann machine, and discuss unsupervised learning algorithms that are suitable to model complicated many-body correlations. We analyze the strengths and weaknesses of conventional and neural-network wave functions by solving various circular quantum-dots systems. Results for up to 90 electrons are presented and particular emphasis is placed on how to efficiently implement these methods on homogeneous and heterogeneous high-performance computing facilities.}

\end{abstract}


\maketitle

\section{Introduction}

Solving the Schrödinger equation for systems of many interacting bosons or fermions is
classified as an NP-hard problem due to the complexity of the required
many-dimensional wave function, resulting in an exponential growth of degrees of freedom. Reducing the dimensionalities of quantum mechanical many-body systems is an important aspect of modern physics, ranging from the development of efficient algorithms for studying many-body systems to exploiting the increase in computing power. To write software that can fully utilize the available resources has long been known to be an important aspect of these endeavors. Despite tremendous progress has been made in this direction, traditional many-particle methods, either quantum mechanical or classical ones, face huge dimensionality problems when applied to studies of systems with many interacting particles. 

Over the last two decades, quantum computing and machine learning have emerged as some of the most promising approaches for studying complex physical systems where several length and energy scales are involved.  Machine learning techniques and in particular neural-network quantum states~\cite{Goodfellow2016} have recently been applied to studies of many-body systems, see for example Refs.~\cite{carleo_solving_2017,carra2021,pfau2019abinitio,calcavecchia_sign_2014,carleo2019,amber2022,Lovato2021,lovato2022}, in various fields of physics and quantum chemistry, with very promising results. In many of these studies, one has obtained results that align well with exact analytical solutions or are in close agreement with state-of-the-art quantum Monte Carlo calculations. 

The variational and diffusion Monte Carlo algorithms are among the most popular and successful methods available for ground-state studies of quantum mechanical systems. They both rely on a suitable ansatz for the ground-state of the system, often dubbed {\it trial wave function}, which is defined in terms of a set of variational parameters whose optimal values are found by minimizing the total energy of the system. Devising flexible and accurate functional forms for the trial wave functions requires prior knowledge and physical intuition about the system under investigation. However, for many systems we do not have this intuition, and as a result it is often difficult to define a good ansatz for the state function. 

According to the universal approximation theorem, a deep neural network can represent any continuous function within a certain error \cite{hornik_multilayer_1989} --- see also Refs.~\cite{Murphy2012,Hastie2009,Bishop2006,Goodfellow2016} for further discussions of deep leaning methods. Since the variational state wave function in principle can take any functional form, it is natural to replace the trial wave function with a neural network and treat it as a machine learning problem. This approach has been successfully implemented in recent works, see for example Refs.~\cite{pfau2019abinitio,carleo_solving_2017,casella2022,Lovato2021,lovato2022}, and forms the motivation for the present study. Here, the neural network of choice was derived from so-called restricted
Boltzmann machines, much inspired by the recent  contributions  by Carleo {\em et al.}, see for example Refs.~\cite{carleo_solving_2017,carleo2019}.  Note that neural-networks representations of variational states are more general, as they do not in principle require prior knowledge on the ground-state wave function, thereby opening the door to systems that have yet to be solved. Particular attention however has to be devoted to the symmetries of the problem, whose inclusion is critical to achieve accurate results~\cite{}.

In this work, we will focus on systems of electrons confined to move in two-dimensional harmonic oscillator systems, so-called quantum dots. These are strongly confined electrons and offer a wide variety of complex and subtle phenomena which pose severe 
challenges to existing many-body methods. Due to their small size, quantum dots are characterized by discrete quantum levels. 
For instance, the ground states of circular dots show similar shell structures and magic numbers as seen for atoms and nuclei. These structures are particularly evident in measurements of the change in electrochemical potential due to the addition of
one extra electron. Here these systems will serve as our test of the applicability of artificial neural network variational states, including restricted Boltzmann Machines.  

The theoretical foundation and the methodology are explained in
section \ref{sec:method}. The subsequent sections present our results with an analysis of computational methods and resources. In the last section we present our conclusions and perspectives for future work.

\section{Method} \label{sec:method}

For any Hamiltonian $\hat{\mathcal{H}}$ and trial wave function $\psi_T$, the variational principle guarantees that the expectation value of the energy $E_T$ is greater than or equal to the true ground state energy $E_0$,
\begin{equation}
E_0 \leq E_T = \frac{\langle\psi_T|\hat{\mathcal{H}}|\psi_T\rangle}{\langle\psi_T|\psi_T\rangle}.
\label{eq:variationalprinciple}
\end{equation}
Thus approximate solutions to the time-independent Schrödinger equation can be obtained by choosing a careful parameterization of the wave function and minimizing the energy $E_T$ with respect to the parameters. Since the integrals representing  $E_T$ are normally  high dimensional, it is most efficient to evaluate them by means of Monte Carlo methods
\begin{equation}
E_T \approx \langle E_L \rangle 
= \frac{1}{n} \sum_{i=1}^n E_L(\bs{R}_i), \ \ \bs{R}_i \sim |\psi_T(\bs{R})|^2.
\label{eq:mcint}
\end{equation}
This involves collecting $n$ samples of configurations and averaging over the so-called local energies
\begin{equation}
E_L(\bs{R}) = \frac{1}{\psi_T(\bs{R})} \hat{\mathcal{H}} \psi_T(\bs{R}).
\label{eq:localenergy}
\end{equation}

We apply the variational Monte Carlo (VMC) method to  various circular quantum dots systems. These are systems of interacting electrons confined to move in   a two-dimensional harmonic oscillator well. The (scaled)\footnote{Natural units are used with energy given in
units of $\hbar$ and length given in units of $\sqrt{\hbar/m}$}
Hamiltonian is given by
\begin{equation}
\hat{\mathcal{H}} = \frac{1}{2}\sum_i \left[-\nabla_i^2 + \omega^2r_i ^2 + \sum_{j\neq i}\frac{1}{r_{ij}}\right],
\label{eq:hamiltonian}
\end{equation}
where $\omega$ is the oscillator frequency, $r_i$ is the distance between electron $i$ and the origin, and $r_{ij}$ is the distance between electrons $i$ and $j$. We will henceforth assume the total number of electrons $N$ to be even and the total spin of the system to be zero.

A simple ansatz can be built starting from the analytical solutions to the non-interacting case. The harmonic oscillator eigenfunctions are given by 
\begin{equation}
    \phi_{m,n}(x,y)\propto e^{-\omega(x^2+y^2)}H_m(\sqrt{\omega}x)H_n(\sqrt{\omega}y),
\label{eq:hermite}
\end{equation}
where $H_n$ are the Hermite polynomials of degree $n$. To constrain the antisymmetry of the many-body wave function, products of the lowest $N/2$ spatial states and the two spin states $\xi_\pm(\sigma)$ are used as a basis for a Slater determinant 
\[
    \psi_{\text{SD}}(\bs{R})=\det \Big[ \Big\{\phi_{m,n}(x_i, y_i)\xi_{k}(\sigma_i) \Big\} \Big],
\]
where $m, n, k$ label the single-particle state, $i$ labels the particle, and $\bs{R}$ contains all coordinates of the $N$ particles. As an aside, we do not include the spin projections $\sigma_i$ as explicit inputs to the wave function as we will describe how to treat them separately in Section II.B. We then define a reference state by pulling the common exponential term out of the determinant and inserting a single variational parameter $\alpha$
\begin{equation}
    \psi_{\text{Ref}}(\boldsymbol{R};\alpha)=e^{-\alpha\omega\sum_i(x_i^2+y_i^2)}\det \Big[ \Big\{ H_m(\sqrt{\omega}x_i)H_n(\sqrt{\omega}y_i) \xi_k(\sigma_i) \Big\} \Big].
\label{eq:ref}
\end{equation}
Correlations among electrons can be handled by a Pad\'e-Jastrow factor
\cite{drummond_jastrow_2004},
\begin{equation}
g(\bs{R}; \beta) = \exp\left(\sum_{i=1}^N\sum_{j>i}^N\frac{a_{ij}r_{ij}}{1+\beta r_{ij}}\right),
\label{eq:padejastrow}
\end{equation}
where $\beta$ is a variational parameter and
\[
    a_{ij}=
    \begin{cases}
        1/3\quad&\text{if } \sigma_i=\sigma_j\\
        1\quad&\text{if } \sigma_i\neq\sigma_j
    \end{cases},
\]
in order for the Kato cusp condition to be satisfied \cite{huang_spin_1998}. The product of the Slater determinant and the Pad\'e-Jastrow factor is commonly named the Slater-Jastrow ansatz,
\begin{equation}
    \psi_{\text{Slater-Jastrow}}(\boldsymbol{R}; \alpha, \beta)=\psi_{\text{Ref}}(\boldsymbol{R};\alpha)
    \times g(\bs{R}; \beta).
\label{eq:slaterjastrow}
\end{equation}

\subsection{Gaussian-binary restricted Boltzmann machine}
\begin{figure}
  \centering
      \includegraphics[width=0.5\textwidth]{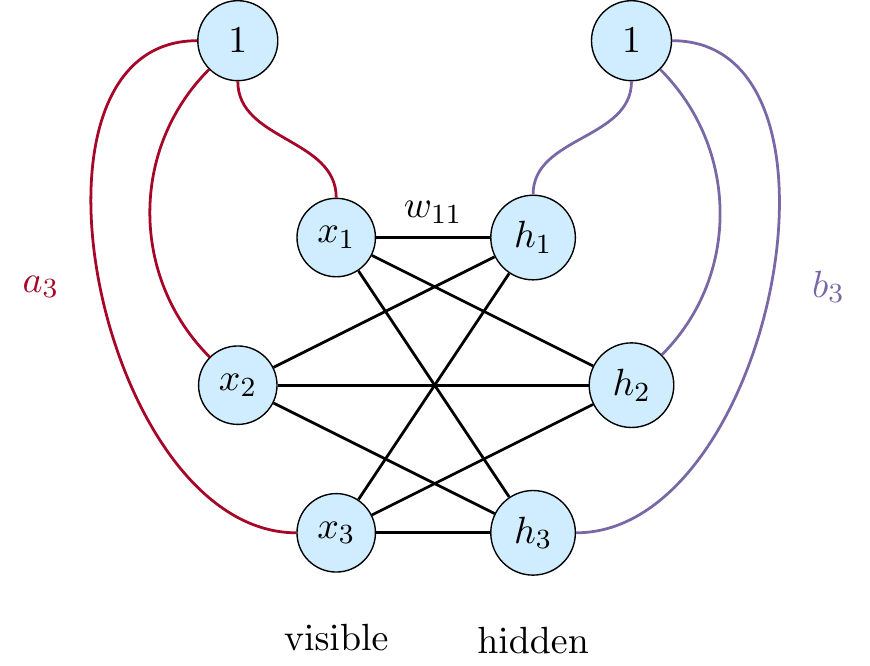}
    \caption{Architecture of a restricted Boltzmann machine. Inter-layer connections between the visible and the hidden layer are represented by the black lines, where, for instance, the line connecting $x_1$ to $h_1$ represents the weight $w_{11}$. The red lines represent the visible biases, where the line going from the bias unit to the visible unit $x_3$ represents the bias weight $a_3$. The purple lines represent the hidden biases, where the line going from the bias unit to the hidden unit $h_3$ represents the bias weight $b_3$.}
    \label{fig:rbmarch}
\end{figure}

There are many possible choices for a machine learning inspired wave function, but using an artificial neural network is natural. Inspired by Ref.~\cite{carleo_solving_2017}, our choice is to start from a restricted Boltzmann machine (RBM) configured for continuous inputs, illustrated in Fig.~\ref{fig:rbmarch}. 
The inputs $\bs{x} \in \mathbb{R}^{2N}$ are the flattened particle positions and interactions between the particles are mediated by $H$ hidden binary nodes. After summing over all the possible values of the hidden nodes, the marginal distribution of the inputs to the Gaussian-binary RBM takes the form
\begin{align}
P(\bs{R};\bs{a},\bs{b},\bs{w})=&\exp\left(-\sum_{i=1}^{2N}\frac{(x_i-a_i)^2}{2\sigma_i^2}\right)\prod_{j=1}^H\left[1+\exp\left(b_j+\sum_{i=1}^{2N}\frac{x_iw_{ij}}{\sigma_i^2}\right)\right].
\label{eq:marginaldist}
\end{align}
Here, $\bs{a} \in \mathbb{R}^{2N}$ and $\bs{b} \in \mathbb{R}^H$ are the bias parameters of the input and hidden nodes, respectively. The weights between the input and hidden nodes are $\bs{w}\in\mathbb{R}^{2N \times H}$, while $\bs{\sigma} \in \mathbb{R}^{2N}$ are the widths of the Gaussian input nodes (not to be confused with the spin projections). It is possible to train these widths by reparameterizing them as $\sigma_i = \exp(s_i)$, but in this work all of the widths were fixed to $\sigma=1/\sqrt{\omega}$ and only the biases and weights are treated as variational parameters. See Appendix \ref{app:deriverbm} for the derivation of the marginal probability.

Notice how the marginal distribution in Eq.~(\ref{eq:marginaldist}) mimics the Gaussian parts of our aforementioned ansätze in Eqs.~(\ref{eq:ref}) and (\ref{eq:slaterjastrow}). Based on such observations, our next step to is construct two corresponding ansätze
\begin{equation}
    \psi_{\text{RBM}}(\bs{R}; \bs{a}, \bs{b}, \bs{w})= P(\bs{R};\bs{a}, \bs{b}, \bs{w}) \times 
    \det \Big[ \Big\{H_m(\sqrt{\omega}x_i)H_n(\sqrt{\omega}y_i)\xi_k(\sigma_i)\Big\}\Big],
\label{eq:rbm}
\end{equation}
and
\begin{equation}
    \psi_{\text{RBM+PJ}}(\bs{R}; \bs{a}, \bs{b}, \bs{w}, \beta)
    =P(\bs{R};\bs{a}, \bs{b}, \bs{w}) \times g(\bs{R}; \beta) \times \det \Big[ \Big\{H_m(\sqrt{\omega}x_i)H_n(\sqrt{\omega}y_i)\xi_k(\sigma_i)\Big\}\Big].
\label{eq:rbmpj}
\end{equation}

The two trial wave functions above apply different levels of physical intuition. 
While $\psi_{\text{RBM}}$ does not contain specific information about the electron-electron interactions, $\psi_{\text{RBM+PJ}}$ contains a correlation factor that explicitly upholds the cusp condition. 
Both ansätze contain knowledge about the required antisymmetry and the Gaussians in the marginal distribution help localize the wave functions to satisfy the boundary conditions far from the oscillator well. Also, as the marginal distribution is positive definite, these ansätze will never collapse into the bosonic state even if the marginal distribution is not symmetric.

\subsection{Code optimization} \label{sec:optimization}

Parallel computing is an important part of our efforts for  developing an efficient VMC solver. However, increasing the available computational resources alone is often not sufficient. One should also consider developing sophisticated algorithms that deliberately minimize the number of floating point operations, cache misses, and communication between parallel processes.

The kinetic energy term of the Schrödinger equation is usually one of
the more computationally expensive parts to compute in terms of computing cycles. It includes
computing, amongst other elements, the Laplacian of the wave function.  The Laplacian term in the expression for the local energy can be written as
\[
\frac{\nabla_i^2\psi_T}{\psi_T}=\nabla_i^2\ln\psi_T + \left(\nabla_i\ln\psi_T\right)^2.
\]
This way of writing the kinetic energy term
is beneficial for two reasons: First, our trial wave function has an
exponential shape, which is taken care of by the log-function.  This is often
the case for many other ansatzes for the trial functions.  Second, this form
allows for separating various elements of the trial wave function.
By writing the trial wave function as a product of
various terms, here $\psi_T=\prod_j\psi_j$, the kinetic energy terms from each particle $i$ can be
written as a sum of their corresponding Laplacians and gradients
\[
    \frac{\nabla_i^2\psi_T}{\psi_T}=\sum_j\nabla_i^2\ln\psi_j+\left(\sum_j\nabla_i\ln\psi_j\right)^2.
\]

Obtaining analytical expressions of the gradient and Laplacian for
all the wave function elements is usually computationally
advantageous. However, in many Monte Carlo studies  they are normally evaluated numerically using automatic differentiation \cite{autodiff2010,autodiff2018}. Nowadays, automatic differentiation  algorithms are employed routinely in  VMC calculations.

The computational complexity of calculating the determinant in the
näive way is proportional to $\mathcal{O}(n^3)$. This calls for a reduction of 
dimensionality as well as efficient evaluations of the Slater determinant.
In this work we do not consider open systems and assume that all single-particle states up to the Fermi level are filled up. We can then split the Slater determinant in a spin-up and a spin-down part
\cite{pfau2019abinitio} without affecting the expectation value of the energy, that is
\[
    \psi_{\text{RBM}}=\det\big[\{\phi_{nm}(\bs{r}_{\uparrow})\xi(\sigma_{\uparrow})\}\big]\times\det\big[\{\phi_{nm}(\bs{r}_{\downarrow})\xi(\sigma_{\downarrow})\}\big].
\]

\begin{figure}[hbtp]
    \centering
    \includegraphics[width=0.7\textwidth]{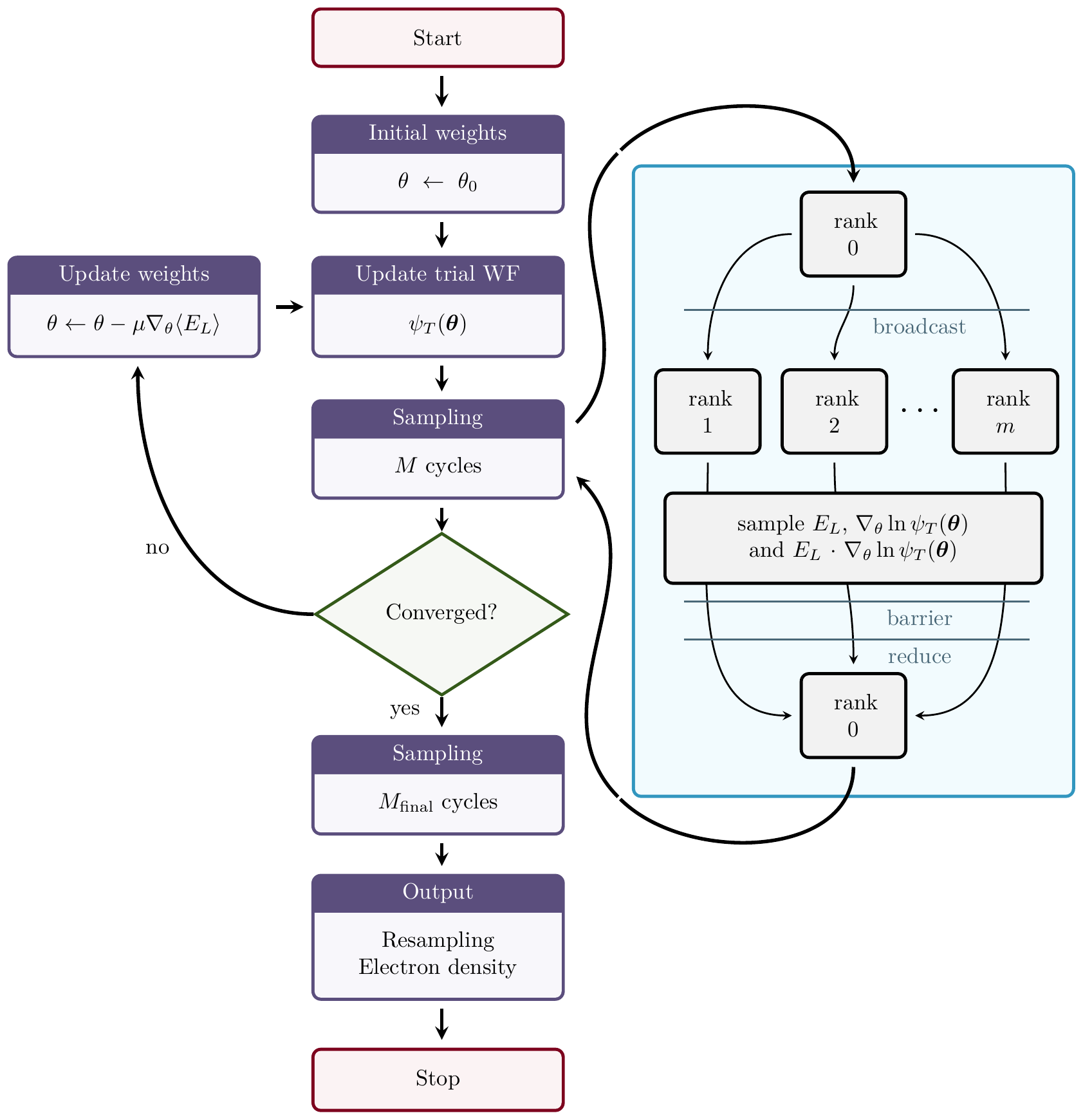}
    \caption{Flow chart of the solver with emphasis on the parallel processing. The sampling is parallelized across $m$ processes, where the trial wave function (WF) is broadcast from process 0. Then $E_L$, $\nabla_{\theta}\ln\psi_T$ and $E_L\cdot\nabla_{\theta}\ln\psi_T$ are sampled independently on each process, and an average is taken after all processes are done with sampling.}
    \label{fig:flowchart}
\end{figure}

The gradient and Laplacian of the logarithm of a determinant with respect to particle $i$ are given by
\[
    \begin{aligned}
    \nabla_i\ln\det=&\sum_j\nabla_id_{ji}d_{ij}^{-1},\\
    \nabla_i^2\ln\det=&\sum_j\nabla_i^2d_{ji}d_{ij}^{-1},
    \end{aligned}
\]
where $d_{ji}$ is element $(j,i)$ of the matrix and $d_{ij}^{-1}$ is
the corresponding element of the inverse matrix
\cite{hammond_monte_1994}.  The general solution requires inversion of
the matrix, which is known to be costly. Fortunately, we can use a
trick to reduce the cost: If we move one particle at the time in our sampling over configurations, this means that we change either the elements of one row or alternatively one column of the Slater
matrix. In this case, there is a simple
relation between the old and the new inverse matrix
\[
    d_{kj}^{-1}=
    \begin{cases}
        \frac{1}{R_i}d_{kj}^{-1}\quad &\text{if } j=i \\
        d_{kj}^{-1}-\frac{S_{ij}}{R_i}d_{ki}^{-1}\quad &\text{if } j\neq i
    \end{cases},
\]
such that the new inverse Slater matrix can be found by a few
operations when the previous is known. Here, $R_i$ is the ratio
between the new and the old determinant and $S_{ij}$ is the cross product
between columns in the new rows and the old matrix,
\[
\begin{aligned}
    R_i=&\sum_j d_{ij}d_{ji}^{-1},\\
    S_{ij}=&\sum_l d_{il}d_{lj}^{-1}.
    \end{aligned}
\]
By using these expressions, the entire Slater determinant matrix is
inverted only once per simulation. We avoid also including spin flips in the simulations. We limit ourselves in this work to systems where we can use these approximations. For systems where all single-particle states up to the Fermi level are filled, the above serves as a useful approximation if the Hamiltonian does not contain spin-dependent terms. If not, every suggested move should include possible spin flips as well.

Like most other Monte Carlo schemes, the algorithm can be split into
smaller individual parts and run efficiently in parallel.  For each
optimization step, the system is sampled independently in several
processes, and the results from all the processes are averaged before
performing the parameter optimization.  In this way, we achieve near
perfect parallelization with message parsing interface (MPI).  A flow
chart of the simulation code can be found in Fig.~\ref{fig:flowchart}.
Notice that the parameters are updated with respect to the gradients
of the expectation value of the local energy. The latter is given by
\[
    \nabla_{\theta}\langle E_L\rangle=2\left(\langle E_L\nabla_{\theta}\ln\psi_T(\bs{\theta})\rangle-\langle E_L\rangle\langle\nabla_{\theta}\ln\psi_T(\bs{\theta})\rangle\right),
\]
as discussed by Ref. \cite{umrigar_energy_2005}, among others.  When
the simulations are run with a burn-in period, each process should
have a burn-in time equal to the burn-in time for a single process.
The theoretical parallelization efficiency is then given by
$(t_{\text{burn-in}}+t_{\text{sample}})/(mt_{\text{burn-in}}+t_{\text{sample}})$
where $m$ is the number of parallel processes, $t_{\text{burn-in}}$ is
the burn-in time and $t_{\text{sample}}$ is the total sampling time.
Additionally, the weight optimization can not be parallelized, but has
a negligible computational cost compared to the sampling. The
communication can also be neglected even with low communication speed.

\section{Results and discussions}

In this section we compare the computational complexity, ground state
energy, energy convergence, contribution from the different
Hamiltonian terms and electron densities for various trial wave
function ansätze. The RBM ansatz consists of a Slater determinant
with Hermite polynomials as the basis, multiplied with the RBM
marginal distribution, as presented in Eq.~\ref{eq:rbm}.  We have not made any attempt to include optimized single-particle state functions through mean-field optimizations like Hartree-Fock theory. The
second ansatz is the RBM ansatz with a correlation factor described in
Eq.~\ref{eq:padejastrow}.  We have also include results obtained
using a traditional Slater-Jastrow ansatz (Eq.~\ref{eq:slaterjastrow}), and as reference we use a plain Slater
determinant (without a correlation factor, Eq.~\ref{eq:ref}).
The diffusion Monte Carlo (DMC) results obtained by
\cite{hogberget_quantum_2013} are our main reference for the ground
state energy.  For quantum dots with two electrons we compare our results with corresponding analytical ones from Ref.~\cite{taut_two_1994}.

\begin{figure}
  \centering
      \includegraphics[width=0.5\textwidth]{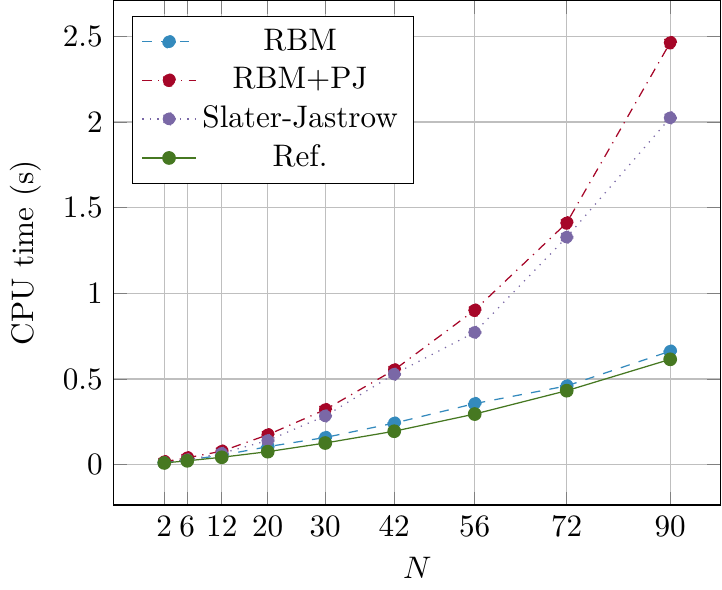}
    \caption{CPU time per iteration as a function of the number of electron $N$ in a quantum dot well. Here we have employed $10^4$ Monte Carlo cycles per iteration (with $10^3$ iterations in total to reach an acceptable statistical error). The number of hidden nodes in the Boltzmann machine was set to $H=6$.}
    \label{fig:cputime}
\end{figure}

In Fig. \ref{fig:cputime} the CPU time per iteration is
plotted as a function of system size for the various ansätze.  The
time per iteration differs strongly for the various ansätze, with the
RBM+PJ and Slater-Jastrow ansätze as the  most
computationally intensive ones.  This is caused by the Padé-Jastrow
factor, which requires all the electron-electron distances at all states.
The execution time is not only longer per iteration, 
but large systems require in general lower
learning rates, and thus significantly more iterations to converge.

The computational overhead associated with parallel processing is
negligible compared to other parts of the code, when ignoring the burn-in steps.  In fact, the
variation in CPU time as a result of noise is much more important than
the variation due to parallel processing.

\begin{table}[h]
  \centering
	\caption{The ground state energy of two-dimensional quantum dots with $N$ electrons and frequency $\omega$. Other references include diffusion Monte Carlo results taken from \cite{hogberget_quantum_2013} and semi-analytical results obtained by \cite{taut_two_1994}. The energy is given in Hartree units, and the numbers in parenthesis are the statistical uncertainties in the last digit. Bold values correspond to the lowest ground-state energy obtained in this work. For abbreviations see the text.\\}
	\begin{tabular}{ccrrrrrr} \hline\hline
		\label{tab:energy2d}
		\makecell{\\ $N$ \\ \phantom{=}} & $\omega$ & RBM & RBM+PJ & Slater-Jastrow & Ref. & \multicolumn{1}{c}{\makecell{DMC \\ (Ref. \cite{hogberget_quantum_2013})}} & {\makecell{Semi-analytical \\ (Ref. \cite{taut_two_1994})}} \\ \hline \\
		2 & 0.1 & 0.46774(5) & \textbf{0.440975(8)} & 0.44129(1) & 0.5279(1) & 0.44079(1) \\ 
		& 1/6 & 0.70389(7) & \textbf{0.666665(1)} & 0.66710(1) & 0.7693(1) & & 2/3 \\
		& 0.28 & 1.07100(6) & \textbf{1.021668(7)} & 1.02192(1) & 1.1388(1) & 1.02164(1) \\
		& 0.5 & 1.72343(7) &  1.659637(6) & \textbf{1.65974(1)} & 1.7983(3) & 1.65977(1)  \\
		& 1.0 & 3.0789(1) & \textbf{2.999587(5)} & 2.99936(1) & 3.1484(3) & 3.00000(1) & 3.0 \\ \hline \\
		
		6 & 0.1 & 3.6971(1) & 3.5700(2) & \textbf{3.5695(1)} & 3.8552(5) & 3.55385(5) \\ 
		& 0.28 & 7.9318(3) & \textbf{7.6203(2)} & 7.6219(1) & 8.0517(9) & 7.60019(6) \\
		& 0.5 & 12.2640(6) & \textbf{11.80494(7)} & 11.8104(2) & 12.2799(9) & 11.78484(6) \\
		& 1.0 & 20.5635(6) & \textbf{20.1773(1)} & 20.1918(2) & 20.697(1) & 20.15932(8) \\ \hline \\
		
		12 & 0.1 & 12.6772(4) & 12.3416(4) & \textbf{12.29962(9)} & 12.9742(9) & 12.26984(8) \\ 
		& 0.28 & 26.389(2) & 25.7266(2) & \textbf{25.7049(4)} & 26.625(2) & 25.63577(9) \\
		& 0.5 & 40.375(1) & \textbf{39.2348(2)} & 39.2421(5) & 40.227(2) & 39.1596(1) \\
		& 1.0 & 67.620(3) & 65.7911(7) & \textbf{65.7026(4)} & 66.744(3) & 65.7001(1) \\ \hline \\
		
		20 & 0.1 & 30.7906(8) &  30.1444(2) & \textbf{30.0403(2)} & 31.253(2) & 29.9779(1) \\ 
		& 0.28 & 63.592(3) & 62.1445(5) & \textbf{62.0755(7)} & 63.681(3) & 61.9268(1) \\
		& 0.5 & 96.356(5) & 94.101(1) & \textbf{94.0433(9)} & 95.755(4) & 93.8752(1) \\
		& 1.0 & 159.428(3) & 156.104(1) & \textbf{155.8900(4)} & 157.904(6) & 155.8822(1) \\ \hline \\
		
		30 & 0.1 & 61.853(2) & 60.774(2) & \textbf{60.585(1)} & 62.449(4) & 60.4205(2) \\ 
		& 0.28 & 126.891(8) & 124.437(2) & \textbf{124.195(2)} & 126.717(5) & 123.9683(2) \\
		& 0.5 & 191.455(9) & 187.488(2) & \textbf{187.325(3)} & 189.977(6) & 187.0426(2) \\
		& 1.0 & 315.364(8) & 308.989(2) & \textbf{308.576(1)} & 311.70(2) & 308.5627(2) \\ \hline \\
		
		42 & 0.1 & 109.767(7) & 108.128(2) & \textbf{107.928(2)} & 110.630(7) & 107.6389(2) \\ 
		& 0.28 & 224.257(9) & 220.588(3) & \textbf{220.224(2)} & 223.837(8) & 219.8426(2) \\
		& 0.5 & 337.43(1) & 331.410(3) & \textbf{331.276(3)} & 335.18(1) & 330.6306(2) \\
		& 1.0 & 553.40(1) & 543.746(3) & \textbf{542.977(2)} & 548.07(2) & 542.9428(8) \\ \hline \\
		
		56 & 0.1 & 179.035(8) & 176.659(2) & \textbf{176.221(1)} & 180.08(1) & 175.9553(7) \\ 
		& 0.28 & 364.52(2) & 359.456(6) & \textbf{358.470(2)} & 363.81(1) & 358.145(2) \\
		& 0.5 & 547.20(3) & 538.666(5) & \textbf{537.841(4)} & 544.12(3) & 537.353(2) \\
		& 1.0 & 894.12(2) & 881.010(5) & \textbf{879.514(3)} & 887.20(5) & 879.3986(6) \\ \hline \\
		
		72 & 0.1 & 274.12(1) & 270.870(3) & \textbf{270.296(3)} & 275.34(2) \\
		& 0.28 & 556.63(2) & 549.899(8) & \textbf{548.315(4)} & 555.45(2) \\
		& 0.5 & 833.85(3) & 822.78(2) & \textbf{821.089(6)} & 829.31(3) \\
		& 1.0 & 1355.37(2) & 1341.54(2) & \textbf{1339.85(1)} & 1349.65(6) \\ \hline \\
		
		90 & 0.1 & 399.84(1) & 395.486(4) & \textbf{394.621(4)} & 401.19(2) \\
		& 0.28 & 809.99(2) & 800.504(6) & \textbf{799.187(5)} & 808.35(2) \\
		& 0.5 & 1211.92(5) & 1198.12(3) & \textbf{1195.025(9)} & 1205.65(4) \\
		& 1.0 & 1973.95(5) & 1948.75(2) & \textbf{1946.27(1)} & 1958.58(5) \\ \hline\hline
	\end{tabular}
\end{table}


The ground state energies of two-dimensional quantum dots of various
sizes and frequencies are presented in table \ref{tab:energy2d}.  The
RBM+PJ and Slater-Jastrow ansätze provide the lowest ground state
energies as expected as the correlation factor does explicit satisfy
the Kato's cusp conditions.  The relative errors with respect to diffusion Monte Carlo (DMC) 
calculations tend to be less than 0.2\% for both methods. The various results obtained with the  RBM+PJ ansatz show that this ansatz
dominates for small quantum dots, but is outperformed by the
Slater-Jastrow ansatz for larger systems.  We suspect this is due to
the fact that the former ansatz is more complex and contains
significantly more parameters than the latter, and has therefore a
hard time finding the global minimum.  The optimization could be
improved with a more sophisticated optimization algorithm.

For low frequency dots ($\omega<0.28$), the RBM produces ground state
energies lower than the reference energy, but fails for large high
frequency dots ($N>6$, $\omega>0.28$).  For low frequency dots the
interaction energy dominates, and the RBM manages to learn the
correlations.  For high frequency dots, the interaction energy is less
important, and the neural network gets redundant.  If not all weights
are set to 0, the neural network will be more a burden for this case.
Additionally, the Gaussian part in the RBM is not identical to the one
in the Hermite functions in the sense that not all coordinates are
multiplied with the same parameter, but rather subtracted with
individual parameters.  Since the interaction part is less important
for these systems, the wave function is closer to the non-interacting
one.  We also observe that the reference values are similar to
corresponding Hartree-Fock energies up to 30 particles
\cite{mariadason_quantum_2018}, which is expected as both approaches
try to find the optimal single Slater determinant without a correlation
factor.

Some simulations were also performed for the RBM and RBM+PJ ansätze with sorted network inputs to enforce anti-symmetry under exchange of two particles, as suggested by Ref.~\cite{saito_method_2018}, among others. Sorted inputs showed promising results, where the ground state energy typically dropped in the fourth or fifth digit. For example, the RBM+PJ ansatz found the ground state energy of a system with $N=30$ electrons and $\omega=0.5$ to be 187.311(1) Hartree. This is lower than the Slater-Jastrow energy. We will investigate this further in a follow-up paper \cite{kim2022}. 

\begin{figure}
  \centering
  \includegraphics[width=0.5\textwidth]{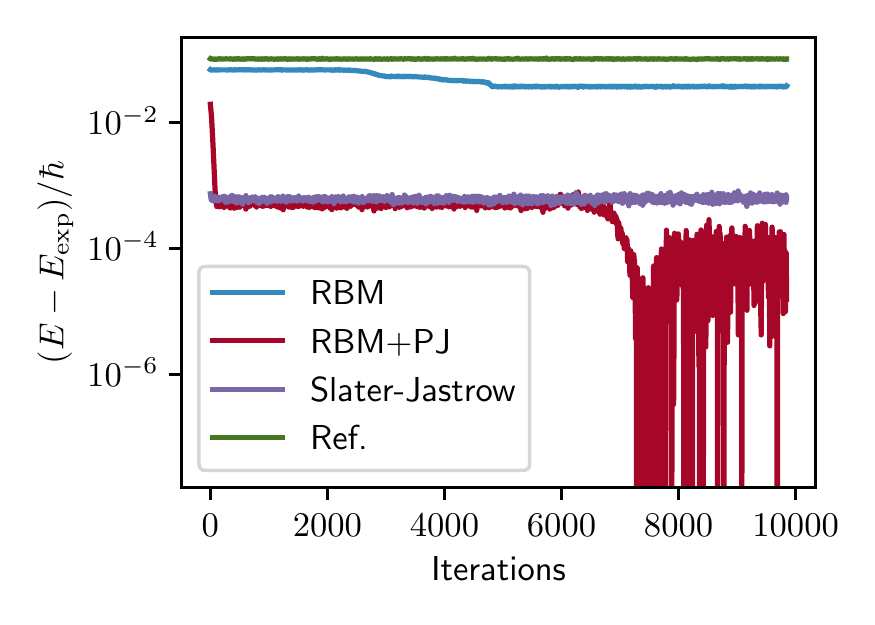}
    \caption{Energy convergence for quantum dots with $N=2$ and $\omega=1/6$. 
    The figure shows how the ground state energy approaches the analytical value (Ref.~\cite{taut_two_1994}) for various ansätze.} 
    \label{fig:energy_convergenceN2}
\end{figure}

\begin{figure}
  \centering
  \includegraphics[width=0.5\textwidth]{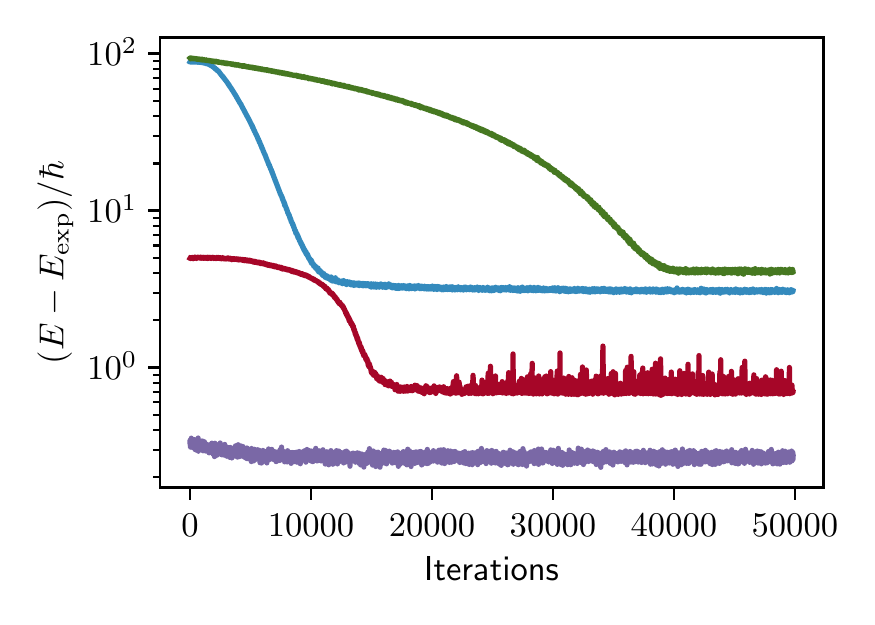}
    \caption{Energy convergence for quantum dots with $N=56$ and $\omega=0.1$. The reference value was obtained by Ref.~\cite{hogberget_quantum_2013} using diffusion Monte Carlo. The labeling of the various results is the same as in the previous figure. }
    \label{fig:energy_convergenceN56}
\end{figure}

For the traditional ansätze that do not contain artificial neural
networks, the learning rate determines to a large extent  how fast the trial wave
function converges towards the true ground state wave function.
However, for the RBM and the RBM+PJ approaches, the results are sensitive to the chosen learning rate values. A too large learning rate can easily
cause exploding energies, and with a too small learning rate the obtained energies with a given trial function might not converge at all.  Our strategy is to find the
highest learning rate that does not lead to exploding energies. In general this requires efficient grid searching methods.  Here, when the energies have
converged, we decrease the learning rate by a factor of 10 and let it
run until it converges again.  Also, the RBMs tend to converge
step-wise, making it hard to know whether or not they have converged.

In Fig.~\ref{fig:energy_convergenceN2}, the convergence of the local energy
is plotted for our four ansätze for $N=2$ electrons. Here the results are compared with analytical calculations for $N=2$.
A similar behavior is shown for  $N=56$ electrons and $\omega=0.1$ in Fig.~\ref{fig:energy_convergenceN56}.  For the 
$N=2$ electrons case, RBM+PJ turns out to be the most accurate ansatz with small absolute errors.  During training,
the Padé-Jastrow parameter is rather constant, but the adjustment of
weights seems important for the model to reach the energy minimum.
For larger systems, the Slater-Jastrow ansatz provides a slightly
lower energy than the RBM+PJ ansatz.  Since the number of inputs for
the RBM increases with number of particles, the network contains 
additional training parameters as we increase the number of electrons.  Therefore,
the networks get more complex for large quantum dots, and they are
naturally harder to train.

\begin{figure}
	\centering
      \includegraphics[width=0.5\textwidth]{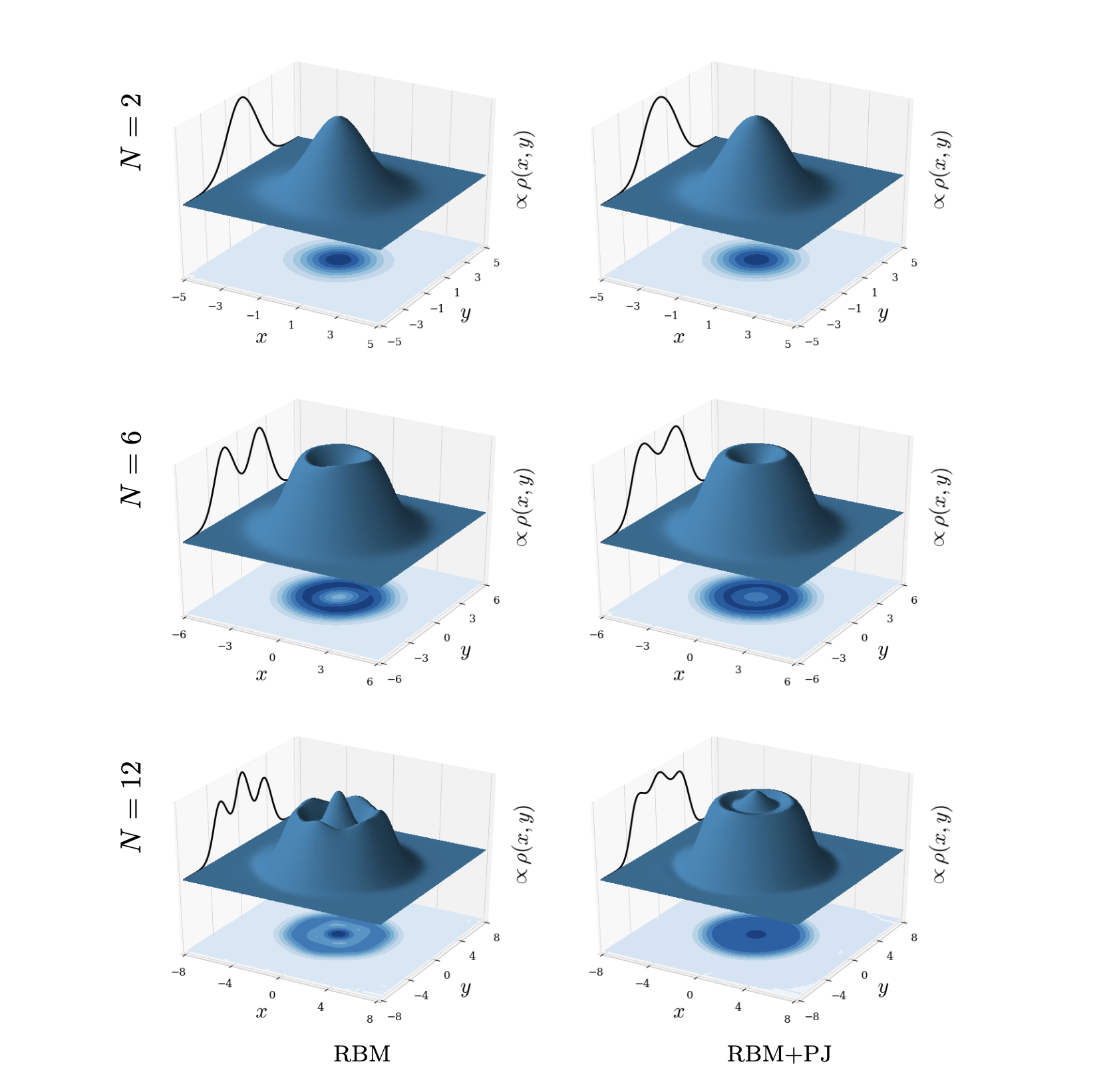}

	\caption{Electron density profiles, $\rho(x,y)$, for two-dimensional quantum dots with frequency $\omega=0.5$ and $N=2$, 6 and 12 electrons seen from the top. The surface plot and the contour plot on the $xy$-plane illustrate the density, and the graph on the $yz$-plane represents the cross-section through $x=0$. They were obtained using RBM (left column) and RBM+PJ (right column), with $M=2^{30}$ Monte Carlo cycles after convergence. The plots are noise-reduced using a Savitzky-Golay filter. For abbreviations and description of the natural units used, see the main text for more details.}
	\label{fig:onebody_w1_1}
\end{figure}

\begin{figure}
	\centering
 \includegraphics[width=0.5\textwidth]{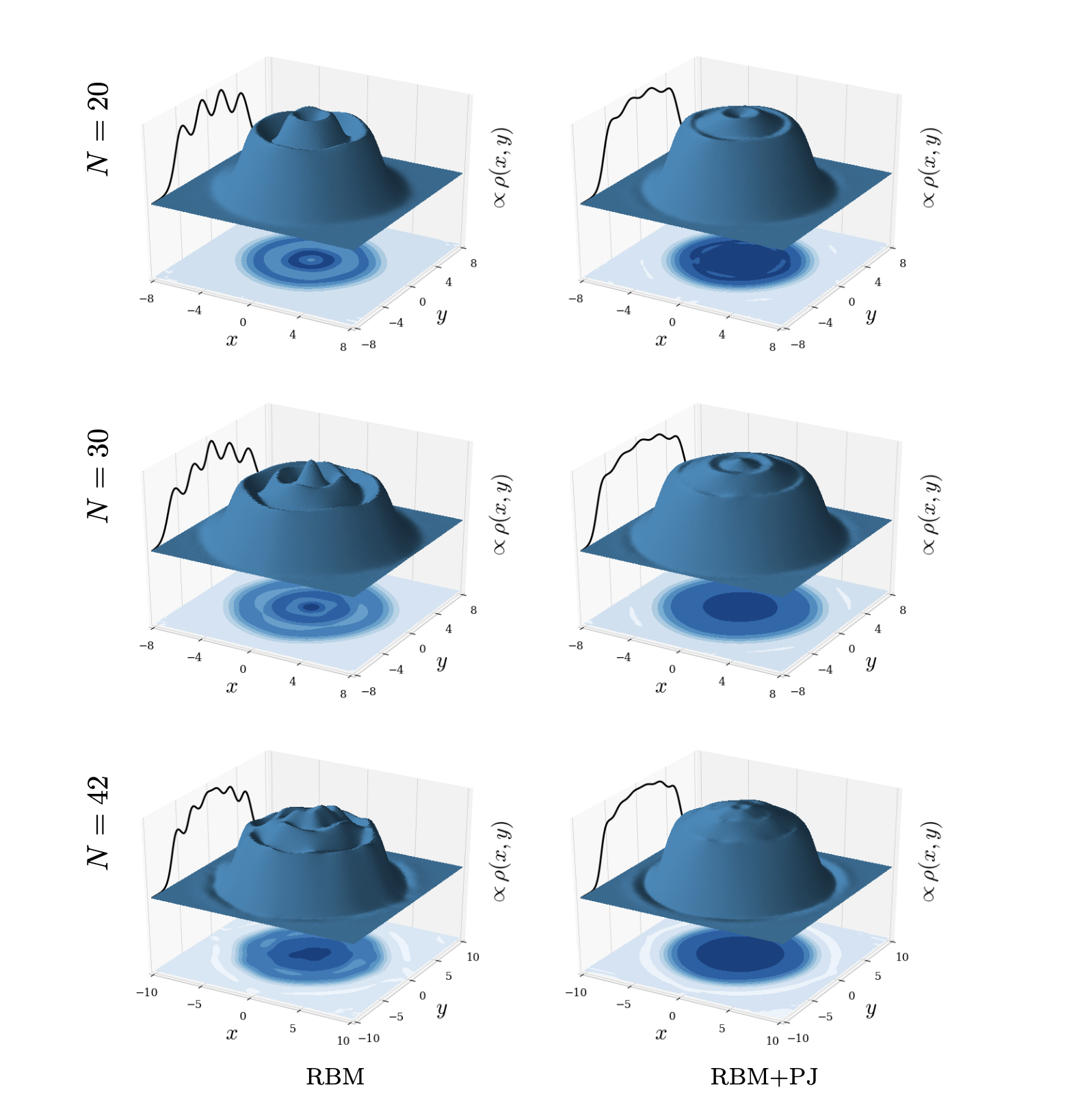}

	\caption{Electron density profiles, $\rho(x,y)$, for two-dimensional quantum dots with frequency $\omega=0.5$ and $N=20$, 30 and 42 electrons seen from the top. The surface plot and the contour plot on the xy-plane illustrate the density, and the graph on the yz-plane represents the cross-section through $x=0$. They were obtained using RBM (left column) and RBM+PJ (right column), with $M=2^{30}$ Monte Carlo cycles after convergence. The plots are noise-reduced using a Savitzky-Golay filter. For abbreviations and description of the natural units used, see main text.}
	\label{fig:onebody_w1_2}
\end{figure}

\begin{figure}
	\centering
 \includegraphics[width=0.5\textwidth]{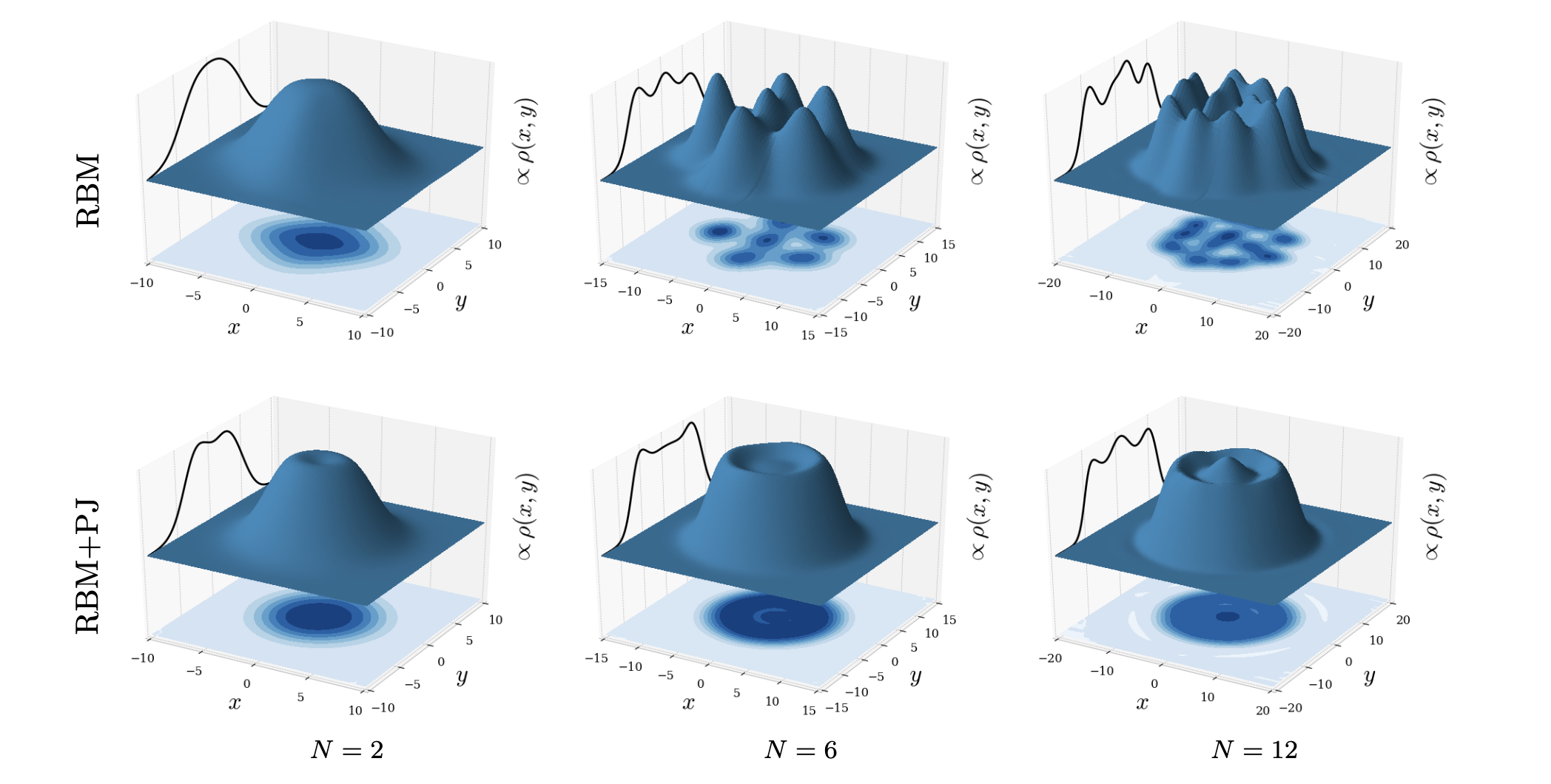}
	\caption{One-body density profile, $\rho(x, y)$, of two-dimensional quantum dots with frequency $\omega=0.1$ and $N=2$, 6 and 12 electrons seen from left to right. The ansätze used are RBM (upper panels) and RBM+PJ (lower panels). The surface plot and the contour plot on the xy-plane illustrate the density, and the graph on the yz-plane represents the cross-section through $x=0$. The surface plots are noise-reduced using a Savitzky-Golay filter. For abbreviations and description of the natural units used, see main text.}
	\label{fig:lowfreq}
\end{figure}

The spatial one-body density plots for $\omega=1.0$ and the RBM and RBM+PJ
ansätze are presented in
Figs.~\ref{fig:onebody_w1_1},\ref{fig:onebody_w1_2}.  The electron densities
have a wave shape for both ansätze, with two nodes for $N=2$, three
nodes for $N=6$ and so on like observed by Refs. \cite{ghosal_incipient_2007,hogberget_quantum_2013}.  The RBM seems to exaggerate the states with
more distinct peaks (higher peaks and lower wave valleys) compared to
the RBM+PJ results.  This can be explained by more localized electrons, which
becomes even more apparent in low frequency dots (see Fig.~\ref{fig:lowfreq}), as the interactions are modelled differently.  The RBM would hardly be able to model the correct electron-electron distances, as the network itself is purely linear.  The
electron density was found to be shape-invariant for high frequency
dots ($\omega>0.28$), with decreasing spatial range as the frequency
increases.

\begin{figure}
	\centering
 \includegraphics[width=0.5\textwidth]{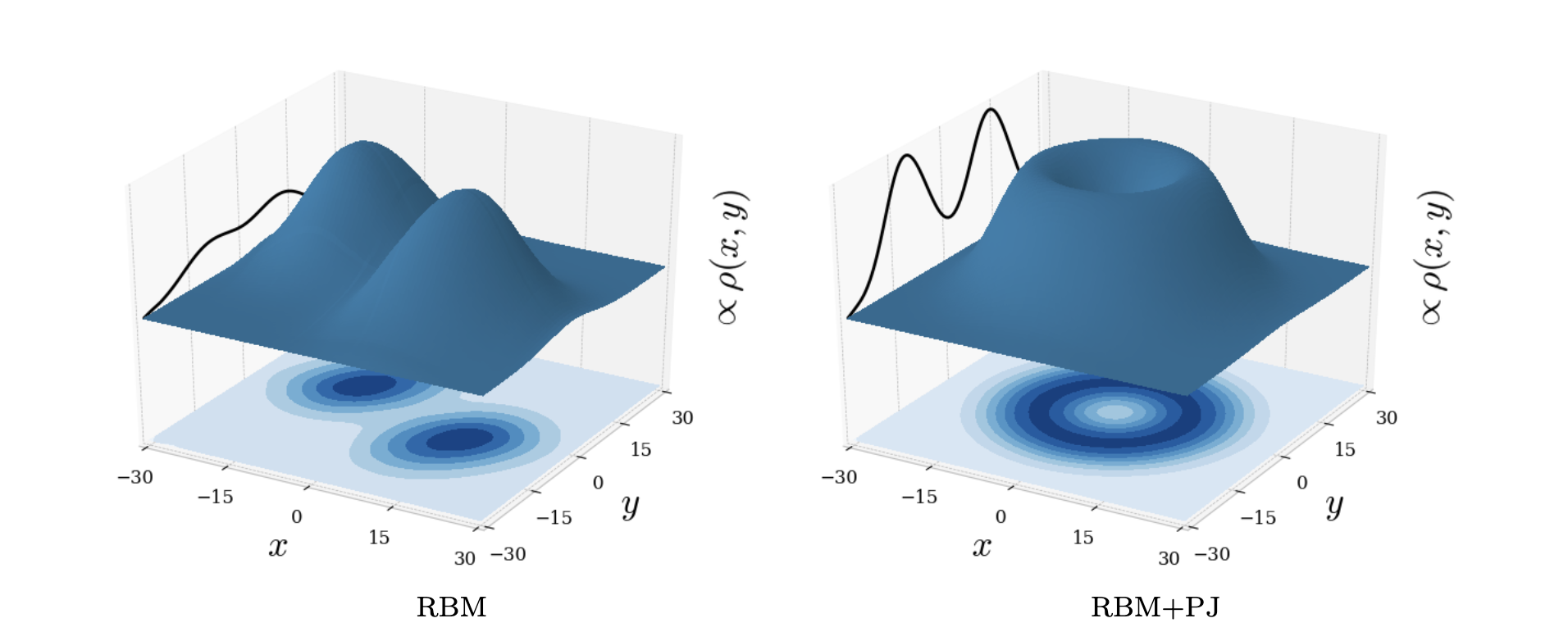}
	\caption{One-body density profile, $\rho(x, y)$, of two-dimensional quantum dots with frequency $\omega=0.01$ and $N=2$ electrons. The ansätze used are RBM (left) and RBM+PJ (right). The surface plot and the contour plot on the xy-plane illustrate the density, and the graph on the yz-plane represents the cross-section through $x=0$. The surface plots are noise-reduced using a Savitzky-Golay filter. For abbreviations and description of the natural units used, see main text.}
	\label{fig:lowfreq2}
\end{figure}

\begin{figure}
    \centering
    \includegraphics[width=0.5\textwidth]{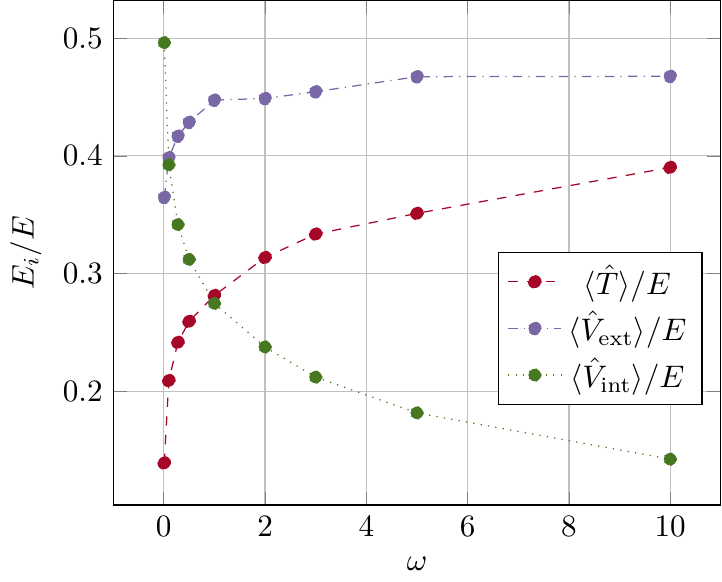}
   \caption{Energy distribution for $N=2$ electrons for the RBM+PJ ansatz, with frequencies ranging from $\omega=0.01$ to $10.0$.}
    \label{fig:energy_distN2}
\end{figure}

\begin{figure}
    \centering
    \includegraphics[width=0.5\textwidth]{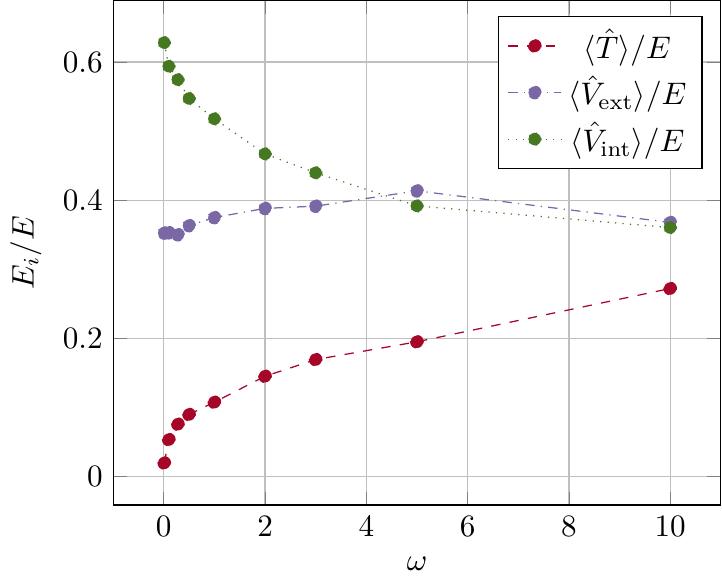}
   \caption{Energy distribution for $N=20$ electrons for the RBM+PJ ansatz, with frequencies ranging from $\omega=0.01$ to $10.0$.}
    \label{fig:energy_distN20}
\end{figure}

When reducing the frequency further, the interaction energy dominates over the kinetic energy and harmonic oscillator energy (see Figs.~\ref{fig:energy_distN2} and \ref{fig:energy_distN20}), and the
electrons naturally become more localized.
The Padé-Jastrow factor solves this by radially localizing the electrons and conserving the circular
symmetry, such that the electrons are confined to specific orbitals.
This can be seen from Fig.~\ref{fig:lowfreq}, where the spatial one-body
density is plotted for low frequency dots ($\omega=0.1$) and system
sizes $N=2$, 6 and 12. Radial localization is also what we would expect from the Hamiltonian, which is strictly circular symmetric.  On the other hand, the RBM seems to localize the electrons both in radial and angular direction, with the number of electrons corresponding to the number of peaks. This is a nonphysical solution to the problem, and shows that the RBM ansatz breaks down for low frequencies.  The RBM+PJ ansatz, on the other hand, confines the electrons in orbitals.  Distinct peaks in radial direction is what is expected from Wigner
crystallization, which we might see indications of with density parameters $r_s\approx 6.7$, $r_s\approx 1.2$ and $r_s=0.3$ respectively for the three system sizes with the RBM+PJ ansatz.  Notice for instance the small ``pit" on top of the $N=2$ plot for RBM+PJ, which is not seen for higher oscillator frequencies.  In Fig.~\ref{fig:lowfreq2}, we have
decreased the one-body density even further to $\omega=0.01$ for $N=2$. As expected, the electrons become even more localized and are clearly showing Wigner crystallization effects with a density parameter of $r_s\approx29$. For the RBM ansatz, the electrons are strongly localized (in all directions) and the electron densities barely overlap. For the RBM+PJ ansatz, we observe strong orbital confinement where the  Wigner crystallization was not the target of this study, but the framework seems capable of a profound study of this phenomenon.

To understand the behavior of the one-body density for various
frequencies, we investigate the expectation values of the kinetic energy ($\langle\hat{T}\rangle$), the harmonic oscillator potential energy
($\langle\hat{V}_{\text{ext}}\rangle$) and the two-body interaction energy
($\langle\hat{V}_{\text{ext}}\rangle$).  The energy distribution is
plotted for the RBM+PJ ansätze for $N=2$ and $N=20$ (Figs.~\ref{fig:energy_distN2} and \ref{fig:energy_distN20}) for frequencies $\omega\in\{0.01$, 0.1, 0.28,
0.5, 1.0, 2.0, 3.0, 5.0 and $10.0\}$.
For large values of the oscillator frequency it is the one-body part of the Hamiltonian which dominates in absolute value (kinetic energy and harmonic oscillator potential energy) compared with the expectation value of the two-body interaction. For such frequencies we notice also that the results can almost be interpreted in terms of the virial theorem. This theorem provides a useful relation between the kinetic and
potential energy \cite{fock_bemerkung_1930}. For circular quantum
dots without the two-electron interaction, the theorem reads
$2\langle\hat{T}\rangle=2\langle\hat{V}_{\text{ext}}\rangle$. From our results we notice that, due to the interaction energy, the ratio between kinetic and harmonic oscillator energies are not exactly equal two for large frequencies. However, for such frequencies the interaction energy plays a less prominent role, resulting thus in a ratio which is close to the ideal value.
When we decrease the frequency however, and the system becomes more dilute with an increase of the mean distance between particles, one notices a somewhat counter intuitive behavior. The expectation value of the kinetic energy and the harmonic oscillator potential energy decrease (and the electrons become more localized as seen in the one-body densities above). However, many-body correlations increase in importance with decreasing frequency. This is reflected in the increased role of the expectation value of the  two-body Coulomb interaction, an effect which is simply due to the infinite range of the Coulomb interaction.  If we were to multiply the Coulomb interaction with a finite range factor, this effect would disappear.   
Figures~\ref{fig:energy_distN2} and \ref{fig:energy_distN20} show this behavior rather clearly.

\section{Computational details}

The quantum dot systems are studied using a general framework for
variational Monte Carlo
simulations (The code is
  available on \url{github.com/evenmn/VMaChine}).  Importance sampling was
used to accelerate the simulations \cite{metropolis_equation_1953}.
To minimize the local energy, we applied the Adam optimizer with
$\beta_1=0.9$ and $\beta_2=0.999$ as suggested by
Ref. \cite{kingma_adam:_2014}.  We use adaptive number of cycles,
starting from $2^{20}$ and then increased to $2^{24}$ for the ten last
iterations after the energy had converged, and further to $2^{30}$ for
the very last iteration to reduce the statistical uncertainty and
noise in electron density.  The particle step length was chosen to get
an acceptance ratio close to 99.5\%, spawning from $10^1$ for the
smallest and weakest systems to $10^{-3}$ for the large and narrow
oscillators.  The optimal learning rate was found by grid searches,
and vary from $10^1$ to $10^{-5}$.  Both the step length and the
learning rate depend strongly on the system and the trial wave function.

For the RBM and RBM+PJ ansätze, only the raw particle positions were input to the RBM. This choice was made for performance reasons, as inputting processed positions like the electron-electron distances would lead to significantly more computations as it would require the processing and increase network complexity.

The Gaussian parameter was initialized to $\alpha=1.0$, which is the
analytical optimum without interaction.  We use Xavier initialization
for the RBM weights, putting they all close to zero
\cite{glorot_understanding_2010}.  The special case with all weights
set to zero corresponds to our reference ansatz with $\alpha=1.0$.
For all the RBMs but the most narrow ones ($\omega=1.0$), the number
of hidden nodes was set to 6, giving 40 to 1272 parameters for the
various system sizes. For $\omega=1.0$, we used $H=12$ hidden nodes to
achieve a lower energy.

All the simulations were run on CPUs.  In total the computational
cost of this project was of the order of $10^6$ CPU hours with largest amount of cycles spent, by obvious reasons, on the largest systems. 

\section{Conclusions and Perspectives}

We found that the RBM ansatz gives a significantly lower ground state
energy than the reference ansatz for low frequency quantum dots. This may  indicate that the RBM manages to capture some of the 
electron-electron correlations.  Based on the one-body density plots,
the RBM found the electrons to be localized both angularly and radially compared to the
ansätze containing a Padé-Jastrow factor, which confine electrons radially only.  For high frequency
dots, the RBM fails in the sense that the obtained ground state energy
is larger than the reference energy.  This can be explained by the
fact that when the interactions get less important, the reference
ansatz is a good guess.  The RBM+PJ ansatz gives energies close to the
DMC energies for small quantum dots.  The ansatz performance needs to
be seen in light of the computational cost, as the ansätze containing
a Padé-Jastrow factor are far more computationally intensive.

From the one-body density and energy distribution plots, it is
apparent that the RBM ansatz is not able to capture the correlations
at the same level as the Padé-Jastrow factor.  Because of the
linearity of the network, it is impossible for it to compute the
distance between the particles, which is crucial to model the
interactions correctly.  One solution to this could be to input the electron-electron distance
into the network, as discussed by \cite{pfau2019abinitio}, but this would increase the computational intensity.  Also,
despite including a Slater determinant, the trial wave function is not
necessarily anti-symmetric when including the an RBM. We performed some simulations where anti-symmetry was forced by sorting the network inputs, which showed promising results in terms of the ground state energy. 
However, although promising results can be obtained, RBMs are  less flexible than general neural networks that make  fewer assumptions about the specific mathematical forms of the trial functions. We have encoded explicitly the anti-symmetry via a Slater determinant. Furthermore, two-body correlations are constructed using a Jastrow factor. An RBM with Gaussian distributions is capable of  capturing the one-body part of the problem, but is less flexible in finding two-body or more complicated many-body correlations. Although the RBM results reported here are promising compared with existing VMC calculations, recent results with neural networks like those presented in for example Refs.~\cite{pfau2019abinitio,casella2022,lovato2022,Lovato2021,carra2021} offer much more flexible and promising research venues for deep learning methods applied to many-body problems. Results obtained with deep neural networks for these systems will be presented in a future work \cite{kim2022}.

\subsection*{Acknowledgements}
AL and BF are supported by the U.S. Department of Energy, Office of Science, Office of Nuclear Physics, under contracts DE-AC02-06CH11357, by the NUCLEI SciDAC program, and the DOE Early Career Research Program.
MHJ is supported by the U.S. Department of Energy,
Office of Science, office of Nuclear Physics under grant
No. DE-SC0021152 and U.S. National Science Foundation Grants
No. PHY-1404159 and PHY-2013047. JK is supported by the U.S. National Science Foundation Grants No. PHY-1404159 and PHY-2013047. EMN is supported by the Norwegian Research Council under grant 287084.

\appendix
\section{Derivation of RBM distributions}
In this appendix, we will derive the marginal and conditional distributions of a Gaussian-binary restricted Boltzmann machine with the system energy
\[
E(\bs{x},\bs{h})=\sum_{i=1}^{2N}\frac{(x_i-a_i)^2}{2\sigma_i^2}-\sum_{j=1}^Hb_jh_j-\sum_{i=1}^{2N}\sum_{j=1}^{H}\frac{x_iw_{ij}h_j}{\sigma_i^2}.
\]
There are $2N$ visible units $x_i$ with related bias weights $a_i$ and $H$ hidden units $h_j$ with related bias weights $b_j$. 
$w_{ij}$ are the weights connecting the visible units to the hidden units. 
The joint probability distribution is given by the Boltzmann distribution
\[
P(\bs{x},\bs{h})=\frac{1}{Z}\exp(-\beta E(\bs{x},\bs{h})),
\]
where $Z$ is the partition function,
\[
Z=\iint d\bs{x}d\bs{h}P(\bs{x},\bs{h}),
\]
and $\beta=1/k_BT$ is the reciprocal temperature that will be fixed to 1. 
As the marginal and conditional distributions are closely related both for the visible and hidden layer, we present the distributions in sections respective for the two layers. 

\subsection{Distributions of visible units} \label{app:deriverbm}
The distributions of the visible units are used to find properties related to the visible units. 
If we recall a restricted Boltzmann machine, the transformation between the visible units and the hidden units is $f_j(\bs{x};\bs{\theta})=b_j+\sum_{i=1}^{2N}w_{ij}x_i/\sigma_i^2$. 
By this expression, we can express the joint probability distribution as
\begin{equation}
\begin{aligned}
P(\bs{x,\bs{h}})&=\frac{1}{Z}\exp\left(-\sum_{i=1}^{2N}\frac{(x_i-a_i)^2}{2\sigma_i^2}+\sum_{j=1}^Hb_jh_j+\sum_{i=1}^{2N}\sum_{j=1}^{H}\frac{x_iw_{ij}h_j}{\sigma_i^2}\right),\\
&=\frac{1}{Z}\exp\left(-\sum_{i=1}^{2N} \frac{(x_i - a_i)^2}{2\sigma^2}+\sum_{j=1}^Hh_jf_j(\bs{x};\bs{\theta})\right).
\end{aligned}
\label{eq:jointvisible}
\end{equation}
The marginal distribution of the visible units is given by the sum over all possible hidden states, $\{\boldsymbol{h}\}\in\{0,1\}$:
\[
P(\bs{x})=\sum_{\{\bs{h}\}} P(\bs{x},\bs{h}),
\]
as the hidden units can take binary values only. 
By inserting the expression of the joint probability distribution from equation \eqref{eq:jointvisible}, we obtain
\[
\begin{aligned}
P(\bs{x})&=\frac{1}{Z}\sum_{\{\bs{h}\}}\exp\left(-\sum_{i=1}^{2N} \frac{(x_i - a_i)^2}{2\sigma^2}+\sum_{j=1}^Hh_jf_j(\bs{x};\bs{\theta})\right),\\
&=\frac{1}{Z}\exp\left(-\sum_{i=1}^{2N}\frac{(x_i - a_i)^2}{2\sigma^2}\right)\times\left( \sum_{\{\bs{h}\}}\prod_{j=1}^H\exp(h_jf_j)\right),\\
&=\frac{1}{Z}\exp\left(-\sum_{i=1}^{2N} \frac{(x_i - a_i)^2}{2\sigma^2}\right)\prod_{j=1}^H\sum_{h_j=0}^1\exp(h_jf_j),\\
&=\frac{1}{Z}\exp\left(-\sum_{i=1}^{2N} \frac{(x_i - a_i)^2}{2\sigma^2}\right) \prod_{j=1}^H \left[1+ \exp(f_j(\bs{x};\bs{\theta}))\right].
\end{aligned}
\]
This is what we will use as the marginal distribution of the visible units. 

\bibliography{References}
\end{document}